# Modeling and Simulation of Regenerative Braking Energy in DC Electric Rail Systems

Mahdiyeh Khodaparastan, *Student Member, IEEE,* Ahmed Mohamed, *Senior Member, IEEE*
Department of Electrical Engineering, Grove School of Engineering, CUNY City College, New York, USA

*Abstract*— Regenerative braking energy is the energy produced by a train during deceleration. When a train decelerates, the motors act as generators and produce electricity. This energy can be fed back to the third rail and consumed by other trains accelerating nearby. If there are no nearby trains, this energy is dumped as heat to avoid over voltage. Regenerative braking energy can be saved by installing energy storage systems (ESS) and reused later when it is needed. To find a suitable design, size and placement of energy storage, a good understanding of this energy is required. The aim of this paper is to model and simulate regenerative braking energy. The dc electric rail transit system model introduced in this paper includes trains, substations and rail systems.

*Keywords*—Electric rail system, regenerative braking energy, simulation and modeling.

## I. INTRODUCTION

Energy efficiency and reducing energy consumption are important challenges in electric rail transit systems. Studies show that up to 40% of the energy supplied to a train can be fed back to the third rail through regenerative braking [1]. Since 1970, electric vehicles with regenerative braking energy capability have been developed [2]. In these trains, during the deceleration, the electromechanical torque of their motors becomes smaller than the load torque. Therefore, the summation of motors torques becomes negative and puts the motors in the generation mode to produce energy. This energy is called regenerative braking energy and can be fed back to the third rail, and absorbed by other nearby accelerating trains. If there are no other neighboring trains, this energy is dumped to a resistor.

Several methods have been proposed for regenerative braking energy recovery, including synchronizing train timetable schedule, reversible substation and installing ESS [3]-[5]. Among these proposed methods, installing ESS is a more popular method.

In addition to reducing energy consumption, using ESS can reduce the peak power demand, which not only benefits the rail transit system but also the distribution utility. ESS may be used to provide services to the main grid, such as shaving peak power demand [6].

In order to select a suitable technology, design, size, and placement of ESS, a good knowledge of regenerative braking energy, including its magnitude, time duration, frequency, etc., is required.

Several studies have been done in the area of modeling and simulation of trains and electric rail systems considering regenerative braking energy [1], [2], [7] and [8]. These studies used a simple model for the train and substation modeling; where the substation is modeled with a DC voltage source with a resistance in series with a diode, and a train is modeled with a current source. In this paper, in order to increase the accuracy of the results from the simulation, more detailed models for the train and substation are presented.

The rest of this paper is organized as follows: Section II provides an overview of the system under study. Section III presents the modeling of the electric vehicle. Simulation and modeling of the substation are presented in section IV. Rails and vehicle movement modeling is presented in section V. Simulation results are presented in section VI, and the conclusion is presented in section VII.

## II. SYSTEM UNDER STUDY

A schematic of a portion of the electric rail transit system is presented in Fig. 1. This portion includes two power supply substations and 3 passenger stations. Trains are running in both directions from west to east and east to west between the passenger stations.
The detailed descriptions about modeling of each component are presented in the following sections.

## III. ELECTRIC RAIL VEHICLE MODELING

There are two main approaches for transient modeling of electric rail vehicles [7].
(1) Cause-effect or forward facing method: In this method, the power consumed by the vehicle is used as an input to determine the speed of the wheel.
(2) Effect-cause or backward facing method. In this method, the speed profile and vehicle properties are used as inputs to determine the input power to the train.

In this paper, the effect-cause approach is used to model the electric rail vehicle. The modeling process is presented in Fig. 2. In this model, the speed of the train is taken as an input, and based on the vehicle dynamic equations represented in (1) to (4), the forces applied to the wheels are calculated.

$$F_T - F_N - F_g - F_a = M_{Metro} \frac{dv}{dt} \quad (1)$$

$$F_N = f_R M_{Metro} g \cos\theta \quad (2)$$

$$F_g = M_{Metro} g \sin\theta \quad (3)$$

$$F_a = \frac{1}{2}C_w A \rho v^2 \quad (4)$$

$$T_w = \frac{F_T \times r}{4n_{cars}} \quad (5)$$

$$\omega_w = \frac{v}{r} \quad (6)$$

Where $F_T$ is the tractive effort, $F_N$ is the force due to friction, $F_a$ and $F_g$ are aerodynamic force and force due to gravity, respectively and $\theta$ is the inclination angle.

From the traction force ($F_T$) calculated in (1), the torque and speed at each axle are calculated using (5) and (6), respectively. The calculated torque and speed are applied to the gearbox. The equations describing the gearbox performance in the motoring and braking conditions are represented in (7) to (10).

$$T_G = \frac{T_w}{\gamma_G} + \frac{B}{\gamma_G} \quad (7)$$

$$\omega_G = \omega_w \gamma_G \quad (8)$$

Where $T_G$ and $\omega_G$ are the gearbox output torque and speed, $\gamma_G$ is the gearbox ratio and $B$ represents the vehicle losses and is calculated using (9).

$$B = T_w(1 - \eta_G) \quad (9)$$

Where $\eta_G$ is the efficiency of the gearbox.

For the braking process (negative torque), (7) can be modified as follow:

$$T_G = \frac{T_w}{\gamma_G} - \frac{B}{\gamma_G} \quad (10)$$

The output torque and speed of the gearbox are applied to the motor drive. The motor drive includes a braking chopper, an inverter, an induction motor and its controller. A schematic of the electric motor drive is presented in Fig. 3.

To model the electric motor drive, a DTC drive available for AC machines in Matlab/Simulink library has been modified and adopted. As presented in Fig. 3, direct torque control (DTC) is used to control the inverter. An overall view of the DTC control method is presented in Fig. 4.

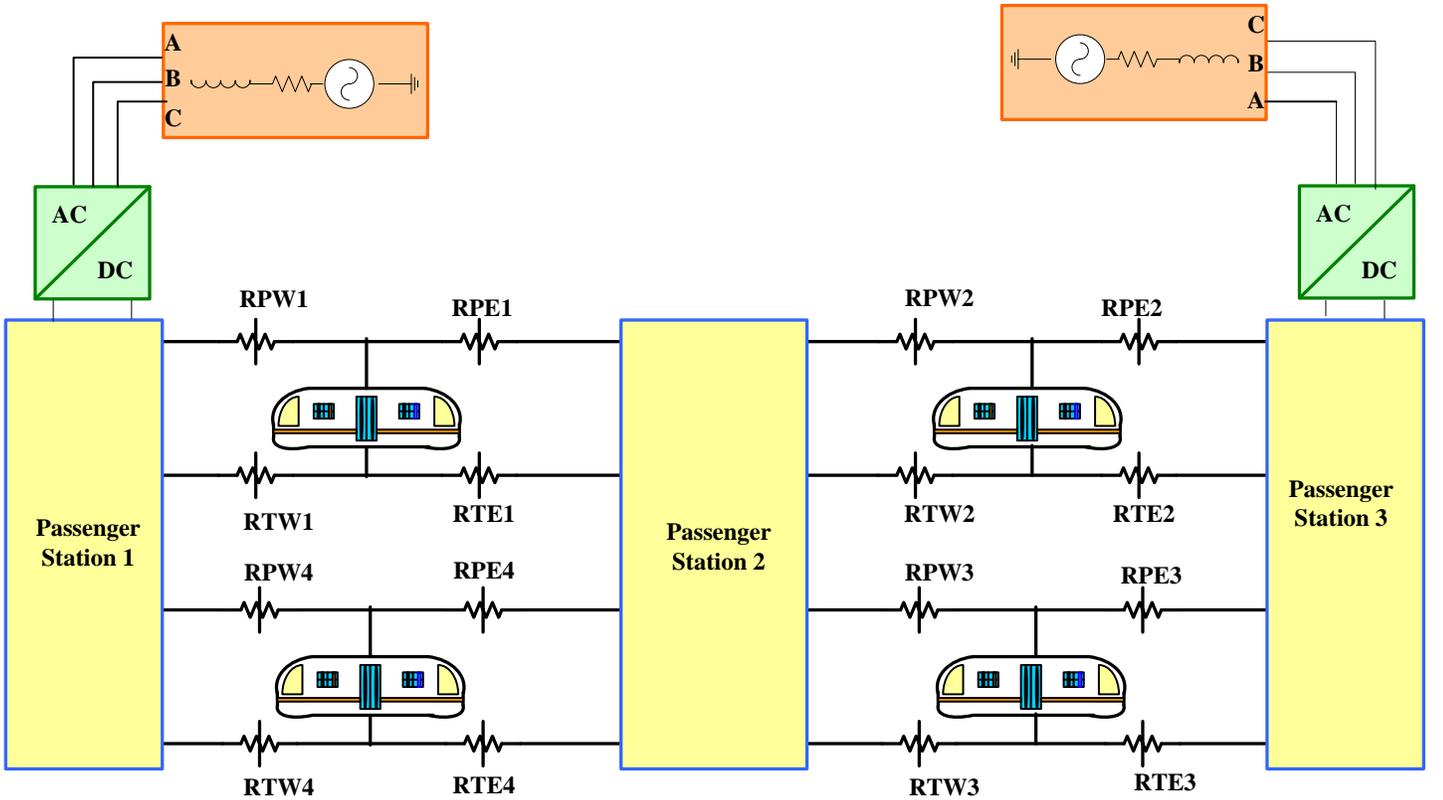

Fig. 1. Overal view of system modeling.

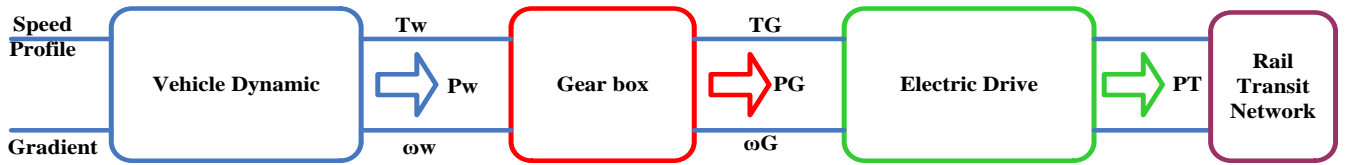

Fig. 2. Overal view of the modeling process.

The torque and flux calculator unit estimates the electric machine torque and stator flux from the terminal voltage and current measurements. The estimated stator flux and electric torque are then controlled directly by comparing them with their respective required values using hysteresis comparators. Then the output of these comparators is used as input to the switching table that produces inverter gate signals.

The torque and the flux references used in this controller are provided by the speed controller unit. A schematic of the speed controller unit is presented in Fig. 5.

The last block in the electric motor drive is the braking chopper. In general, injecting regenerative braking energy to the third rail will increase the voltage level of the third rail. If this increase goes beyond a certain level, protection devices will act and do not let this energy get injected into the rail. The rest of this energy will be dumped in onboard resistance through the braking chopper. A schematic view of the braking chopper is presented in Fig. 6. The chopper circuit consists of a resistor bank and an IGBT switch controlled by hysteresis controller. In this controller, the DC bus voltage is monitored and if the voltage goes above a pre-specified value (upper limit activation voltage), the IGBT switch turns on. Then, the excess power on the line will be dissipated in the resistor bank. The switch is turned off when the voltage drops to the lower voltage limit (lower limit shut down voltage) as presented in Fig. 7.

The parameters used for electric vehicle modeling are presented in Table I.

Table I. Modeling Parameters of the Vehicle

| Parameters | Unit | Value |
|---|---|---|
| The vehicle mass for one car ($M_{Metro}$) | T | 38 |
| The rolling resistance factor($f_R$) | - | 0.002 |
| The Gravity (g) | N/Kg | 9.81 |
| The rail slop($\alpha$) | - | 0 |
| The drag coefficient($C_W$) | - | 0.5 |
| The air density ($\rho$) | kg/m$^3$ | 1.225 |
| The front area of the train (A) | m$^2$ | 9 |
| The wheel radius (r) | M | 0.432 |
| The number of cars ($n_{cars}$) | - | 10 |
| the gearbox ratio ($\gamma_G$) | - | 6.64 |
| efficiency of gearbox ($\eta_G$) | - | 0.96 |
| Chopper resistance | Ω | 2 |
| Kp and Ki for speed Controller | - | [30, 100] |
| Torque Limit | N.m | [-2000, 2000] |
| acceleration Limit | Mph/s | [-3, 3.5] |

## IV. SUBSTATION MODELING

A typical DC transit substation consists of a voltage transformation stage that steps down the medium voltage to a lower voltage level, followed by an AC/DC rectification stage that provides DC power to the third rail. There are also traction network protection devices both at the AC and DC sides to prevent personnel injuries, equipment damage and interference to other users.

The electric power supply substation has been modeled by two parallel circuits as shown in Fig. 8. One of the circuits consists of a three-phase Δ/Δ transformer in series with an

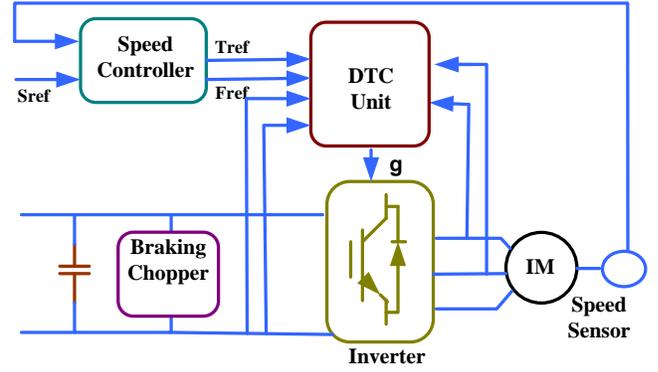

Fig. 3. Schematic of Motor Drive.

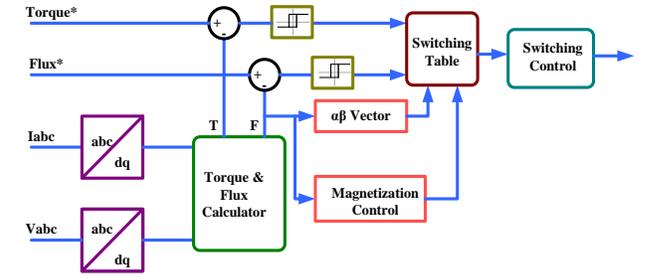

Fig. 4. Block diagram of the direct torque controller (DTC).

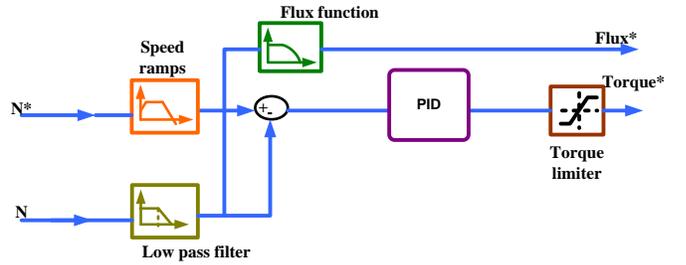

Fig. 5. Block diagram of the speed controller.

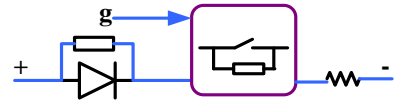

Fig. 6. Block diagram of the braking chopper.

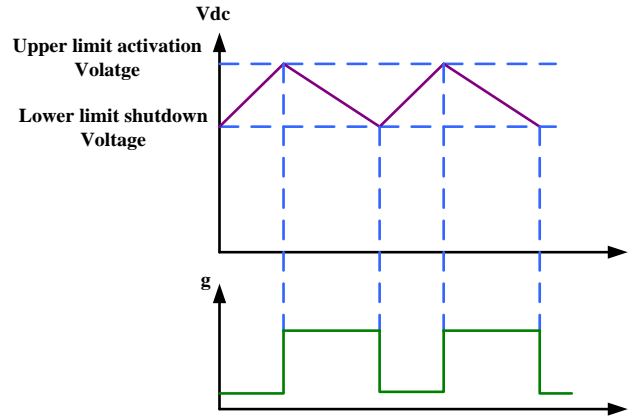

Fig. 7. Control method of braking chopper switch.

AC/DC converter. The other circuit consists of three phase Y/Δ transformer. An auxiliary transformer is also modeled, which provides supply to the loads, such as elevators, escalator, ventilation systems and lighting systems [9].

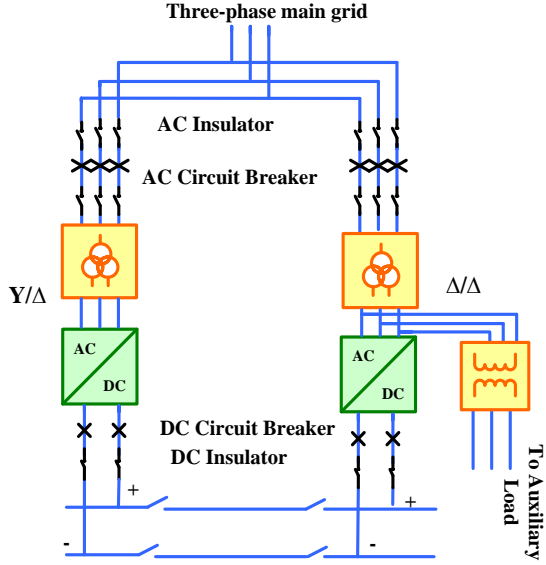

Fig. 8. Block diagram of substation model.

## V. ELECTRIC RAIL SYSTEM MODELING

To model an electric train moving on the rail, traction and power rails are modeled by variable resistances located on the west and east sides of the train, as shown in Fig. 1. The value of these variable resistances changes based on the train position. Fig. 9 shows a block diagram of the calculating resistance at each time step [10].

The equations related to the variable resistance of each section of the westbound rails are presented in (11) to (18). The resistance of the eastbound rails can be calculated in a similar way.

$$R_{WP1} = (p - x_{p1}) \times R_{Ppu,rail} \tag{11}$$

$$R_{WT1} = (p - x_{p1}) \times R_{Tpu,rail} \tag{12}$$

$$R_{EP1} = (x_{p2} - x_{p1} - p) \times R_{Ppu,rail} \tag{13}$$

$$R_{ET1} = (x_{p2} - x_{p1} - p) \times R_{Tpu,rail} \tag{14}$$

$$R_{WP2} = (p - x_{p2}) \times R_{Ppu,rail} \tag{15}$$

$$R_{WT2} = (p - x_{p2}) \times R_{Tpu,rail} \tag{16}$$

$$R_{EP2} = (x_{p3} - x_{p2} - p) \times R_{Ppu,rail} \tag{17}$$

$$R_{ET2} = (x_{p3} - x_{p2} - p) \times R_{Tpu,rail} \tag{18}$$

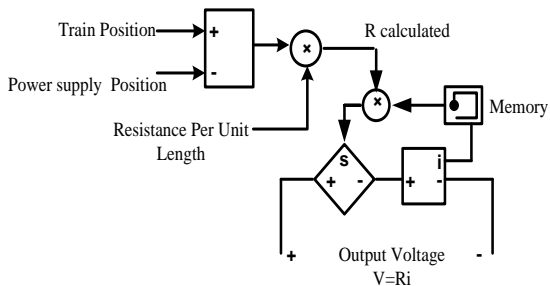

Fig. 9. Block diagram of substation model.

Where $p$ is the position of the train, $x_{pi}$ is the position of the $i_{th}$ passenger station, $R_{Ppu,rail}$ and $R_{Tpu,rail}$ are per-unit length resistance of the power and traction rails, respectively. The resistances on the left-hand side of the equations are the resistance of each section of rails as shown in Fig. 1.

## VI. SIMULATION RESULTS

In order to quantify the energy regenerated during deceleration of the train, the results of a train running between passenger station one and two in westbound rail is presented. The speed profile, current and one of the motor outputs (including stator current, speed, torque and voltage) are presented in Figs. 10 to 12. As can be seen from Fig. 12, the motor speed and torque are following their reference values.

The electric vehicle power and related energy are presented in Figs. 13 and 14. A change in the energy level for one electric vehicle running from the first passenger station to the second one during acceleration (0s-37.4s) is approximately 74.61 MW.s (20.72 kWh) and the energy during deceleration (37.4s-63s) is approximately 57.57 MW.s (16kWh). As can be concluded, almost 78% of the energy used during acceleration can theoretically be regenerated by train during deceleration.

Three different speed profiles are also fed into the simulation model and the results for their current and energy are presented in Fig.15. For the first case, the change of total energy during acceleration (0 s-21.3s) is about 13 kWh and the change of total energy during deceleration (55s-72 s) is about 10kWh (about 76% of energy consumed for train acceleration).

Similarly, in the second case, almost 13kWh is consumed during acceleration while almost 9kWh is produced during deceleration. While in the third case, change in the energy level during acceleration phase is about 12 kWh and the change in energy level during deceleration is about 10 kWh (about 83% energy consumed during acceleration).

By looking precisely into the speed and current profiles, it can be concluded that the magnitude of regenerative braking energy can vary based on the rate of deceleration. Increase in the deceleration rate results in the higher current magnitude. For instance, compare the magnitude of train current case two (about -4000A) in fig.15 where the speed has a moderate deceleration rate with the magnitude of current in case 3 (about -9000 A) in Fig.15 where the speed has a stipe deceleration rate.

It is worthy to mention that not all this energy can practically be recuperated. As illustrated in Fig. 11, the regenerative current is released with the high magnitude in a very short time interval at the beginning of deceleration. If this current is let to flow to the third rail, the voltage will tend to rise to a high value and before it reaches its maximum, the protection devices will disconnect the train, which will dump the energy into its board resistance.

The magnitude of the current that is created during deceleration depends on the deceleration rate and maximum speed at the beginning of deceleration. In the other hand, the amount of energy that can be injected to the third rail depends on the receptivity of system. If there are other loads, such as nearby accelerating trains available in the system and the system has higher overvoltage protection setting, more regenerative braking energy can be injected to the third rail.

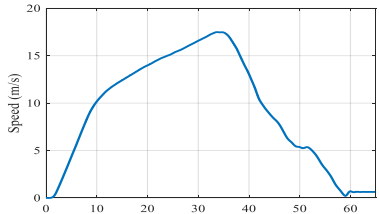

Fig. 10. Speed profile.

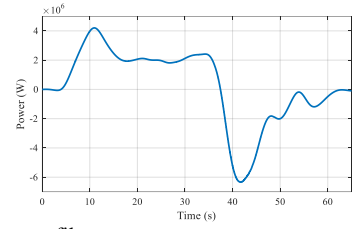

Fig. 13. Train power profile.

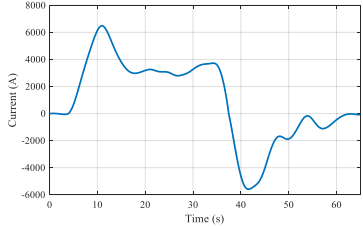

Fig. 11. Train current profile.

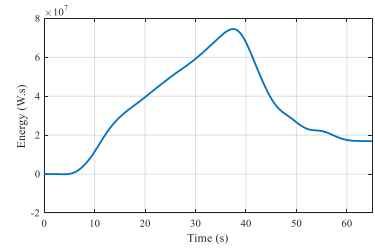

Fig. 14. Train energy profile.

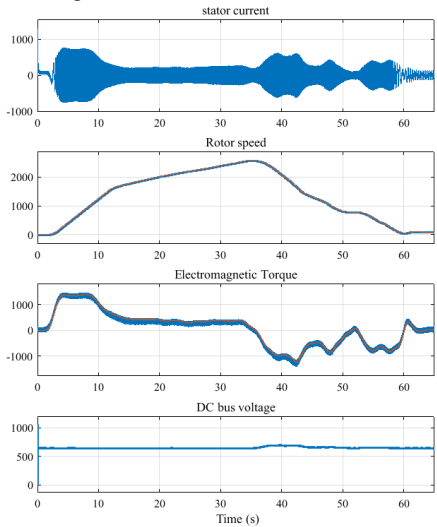

Fig. 12. Motor output.

## VII. Conclusion

Energy saving and increasing energy efficiency are important challenges in electric rail transit systems. One solution is the recuperation of regenerative braking energy by installing energy storage systems. In order to design, size and determine the suitable placement of energy storage systems, quantifying the available regenerative braking energy is imperative. This paper has presented a detailed modeling and simulation approach for regenerative braking energy in a DC electric rail system. The system under study includes two power supply substations, electric rail vehicles and a rail system. The modeling details along with related equations are presented for each component. Results show that the proposed models can provide a reliable simulation tool for regenerative braking energy.

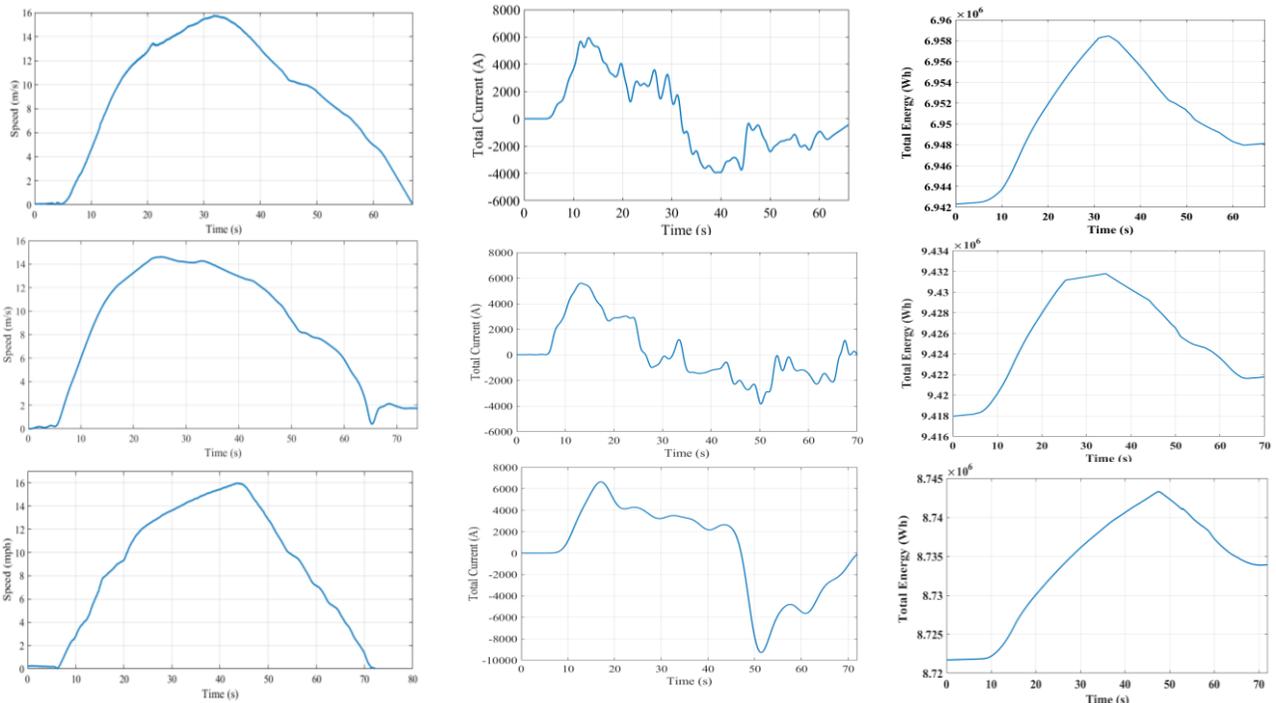

Fig.15. Simulation result for different speed profile.